\documentclass[aps,prl,nopacs,lettersize]{revtex4}

\usepackage{color,graphicx}
\usepackage{times}
\usepackage[sort&compress]{natbib}
\newcommand{\citenum}{\cite}
\begin{document}

\title{Spontaneous vortices in the formation of Bose-Einstein condensates}

\author{C.~N.~Weiler$^1$, T.~W.~Neely$^1$, D.~R.~Scherer$^1$,
A.~S.~Bradley$^2$, M.~J.~Davis$^2$, \& B.~P.~Anderson$^1$}

\affiliation{$^1$College of Optical Sciences, University of Arizona,
Tucson, AZ 85721, USA\\
$^2$ARC Centre of Excellence for Quantum-Atom Optics,
 School of Physical Sciences, University of Queensland, Brisbane,
 Queensland 4072, Australia
}

\begin{abstract}
\textbf{
Phase transitions are ubiquitous in nature, ranging from protein folding and denaturisation, to the superconductor-insulator quantum phase transition, to the decoupling of forces in the early universe.  Remarkably, phase transitions can be arranged into universality classes, where systems having unrelated microscopic physics exhibit identical scaling behaviour near the critical point. Here we present an experimental and theoretical study of the Bose-Einstein condensation phase transition of an atomic gas, focusing on one prominent universal element of phase transition dynamics: the spontaneous formation of topological defects during a quench through the transition \cite{Kibble1976a,Zurek1985a,Zurek1996a}.
While the microscopic dynamics of defect formation in phase transitions are generally difficult to investigate, particularly for superfluid phase transitions \cite{Hendry1994a,Ruutu1996a,Bauerle1996a,Dodd1998a}, Bose-Einstein condensates (BECs) offer unique experimental and theoretical opportunities for probing such details.  Although spontaneously formed vortices in the condensation transition have been previously predicted to occur \cite{Anglin1999a,Svistunov2001a}, our results encompass the first experimental observations and statistical characterisation of spontaneous vortex formation in the condensation transition.  Using microscopic theories \cite{Stoof1999a,Davis2001a,Davis2001b,Gardiner2002a,Gardiner2003a,Blakie2005a,Davis2006a,Bradley2007a} that incorporate atomic interactions and quantum and thermal fluctuations of a finite-temperature Bose gas, we simulate condensation and observe vortex formation in close quantitative agreement with our experimental results.  Our studies provide further understanding of the development of coherence in superfluids, and may allow for direct investigation of universal phase-transition dynamics.
}
\end{abstract}

\maketitle

\begin{figure}
\includegraphics{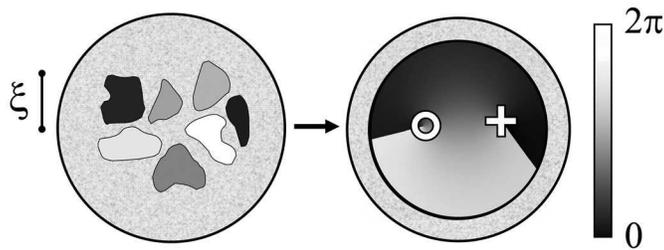}
\caption{\label{suppfig} \textbf{$|$~~Schematic of spontaneous vortex formation.} (Left) As a thermal gas (mottled grey shade) is cooled through the BEC transition, isolated coherent regions of approximate size $\xi$ and
unpredictable phase may form \cite{Anglin1999a,Svistunov2001a}.  Quantum phase ranges from $0$ to $2\pi$, represented here by shades of grey as indicated by the gradient bar at the right. (Right) Initial coherent regions eventually merge to form a single BEC (continuous greyscale region), potentially forming quantised vortices.   Here, a positive (negative) vortex is labeled with a cross (circle), with the phase winding direction corresponding to the direction of superfluid flow and phase gradient around the vortex core.}
\end{figure}

Spontaneous vortex formation in superfluids is intimately connected to superfluid growth.  In one model, illustrated in Fig.~\ref{suppfig}, isolated superfluid regions of characteristic size $\xi$ independently form as the system nears the critical point of the phase transition.  These regions with random relative phases merge together during the transition, leading to a continuous phase \emph{gradient} in the merged fluid. Due to wave function continuity requirements, the merging process may trap phase loops of $2\pi$ if the merging regions have suitable relative phases, as illustrated in Fig.~\ref{suppfig}. The superfluid density at the center of these $2\pi$ phase loops is topologically constrained to be zero, resulting in the formation of a \textit{quantised vortex}; the absence of superfluid at the vortex core may be viewed as arising from destructive interference between merging regions.  Although cast here in the context of superfluid growth, spontaneous topological defect formation is a fundamental component of the Kibble-Zurek (KZ) mechanism \cite{Kibble1976a,Zurek1985a,Zurek1996a}.  Based on universality classes for second-order phase transitions, the KZ mechanism provides a prescription for estimating a correlation length $\xi$ and hence the density of defects, proportional to $1/\xi^2$, that may form. For a continuous phase transition that proceeds quasistatically, $\xi$ diverges at the critical point and therefore no defects are expected to form.  However, in the KZ mechanism the phase transition occurs over a finite time, and the system falls out of equilibrium when the thermalisation (or relaxation) rate drops below a quench rate $1/\tau_Q$.  At this point $\xi$ is frozen in, essentially remaining constant through the critical point.  A principle result is that faster quenches lead to an earlier freeze-in time, and hence smaller values of $\xi$ and higher defect densities.

The KZ mechanism is appealing due to its potential for characterizing a wide variety of phase transitions, irrespective of the microscopic processes involved. A model of condensation in a homogeneous Bose gas describing the transition from a weak-turbulent (kinetic) stage to strong-turbulent (coherent) state has been proposed by Svistunov and co-workers \cite{Svistunov1991,Kagan1992,Kagan1994,Kagan1997,Svistunov2001a}.  In this scenario, as energy is removed from a system, the low-energy atomic field modes become macroscopically occupied. Destructive interference between these essentially classical modes leads to nodes in the field, which appear as lines of zero atomic density.  Subsequently, a \emph{quasi-condensate} having local coherence but no long-range coherence grows around the lines of zero density, which simultaneously evolve into well-structured vortex cores.  Eventually the superfluid relaxes into equilibrium and a true condensate with global phase coherence is achieved.  Berloff and Svistunov numerically demonstrated the validity of this scenario for the \emph{homogeneous} Bose gas in simulations of the Gross-Pitaevskii equation \cite{Berloff2002a}.  Our work involves an experimental and theoretical exploration of similar phenomena in the condensation of trapped gases. Our approach also has the potential to investigate the relationship of the KZ mechanism to BEC phase transition dynamics.

In previous work we demonstrated that vortices can form during the \emph{controlled} merging of three independent BECs with uncorrelated phases \cite{Anderson2006}, an analog of the KZ mechanism.   Here we study vortex formation by evaporatively cooling an atomic gas through the BEC phase transition in a single axially symmetric oblate harmonic trap (see Methods).  In order to probe condensate growth dynamics under varying cooling conditions, we utilise two temperature quenches: Quench A uses a 6-s radio-frequency (rf) evaporative cooling ramp, and Quench B uses a sudden rf jump to a final value.  Plots of temperature and condensate number versus time for both quenches are shown in Fig.~\ref{fig1}a.  Following Anglin and Zurek \cite{Anglin1999a} (see Supplementary Information), we estimate a correlation length of ${\xi}\approx 0.6$  $\mu$m near the critical point for both quenches.  Because $\xi$ is about a factor of 5 \emph{smaller} than our radial harmonic oscillator length $a_r \sim 3.8$ $\mu$m characterising a condensate radius for small atom numbers, we would not expect global phase coherence just below the critical temperature, suggesting that spontaneous vortex formation could be observable in our experiments.

\begin{figure}
\includegraphics{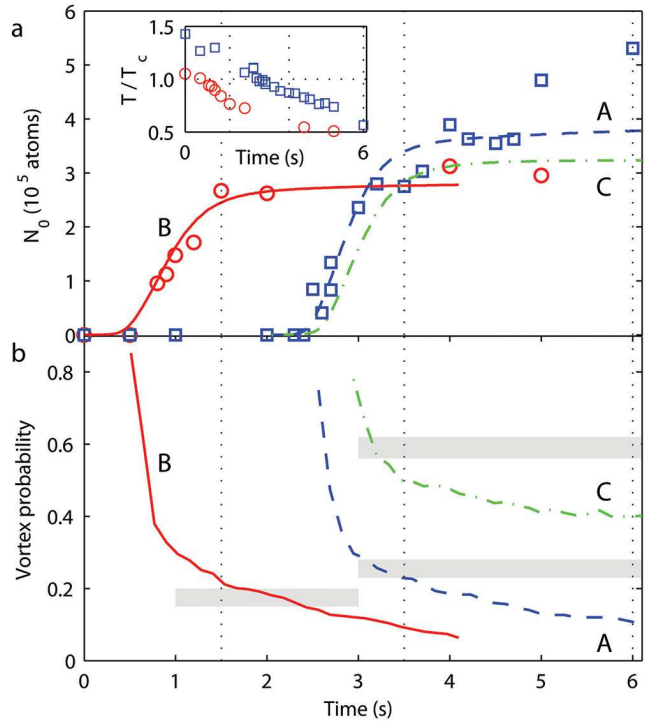}
\caption{\label{fig1}\textbf{$|$~~Condensate formation and vorticity.}  \textbf{a},  Condensate number $N_0$ versus time.  Blue squares (red circles) indicate experimental data for Quench A (B), and lines indicate corresponding theoretical simulations.  The green dot-dashed line is the theoretical result for the toroidal trap (Quench C).  Vertical dotted lines indicate the observation times for which experimental statistics are generated.  Inset: experimentally measured temperatures for Quenches A and B ($\tau_Q \sim$ 7 s and 5 s, respectively).  \textbf{b}, The probability of finding at least one vortex passing through the $z=0$ plane plotted for all three simulated quenches.  Grey bars indicate the experimental measurement range for each data set.}
\end{figure}

\begin{figure}
\includegraphics{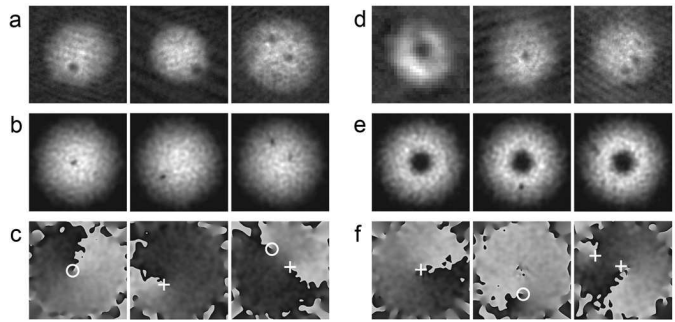}
\caption{\label{fig2}\textbf{$|$~~Vortices in the harmonic and toroidal traps.}    \textbf{a}, 200-$\mu$m-square expansion images of BECs created in a harmonic trap, showing single vortices (left, centre) and two vortices (right).  \textbf{b, c}, Sample simulation results from Quench B, showing in-trap integrated column densities along $z$ (in \textbf{b}) and associated phase profiles in the $z=0$ plane (in \textbf{c}), with vortices indicated by crosses and circles at $\pm 2\pi$ phase windings.  \textbf{d}, Left image: 70-$\mu$m-square phase-contrast experimental image of a BEC in the toroidal trap.  Remaining images: vortices in 200-$\mu$m-square expansion images of BECs created in the toroidal trap.  \textbf{e, f},   Simulations of BEC growth in the toroidal trap show vortices (as in \textbf{b},\textbf{c}) and persistent currents.}
\end{figure}

To look for vortices, we suddenly remove the trapping potential after the 6-s evaporative cooling ramp of Quench A or $\sim$1.5 s after the rf jump for Quench B.  Each BEC ballistically expands and is then imaged along the vertical direction (the $z$ axis), which coincides with the trap's symmetry axis.  Vortex cores well-aligned with the $z$ axis appear as holes in the column-density distribution, as shown in Fig.~\ref{fig2}a.   We emphasise that our procedure does not impart net angular momentum to the atomic cloud, such as through phase engineering \cite{Matthews1999} or stirring \cite{Madison2000}; our observations thus represent a new regime for the study of quantised vortex nucleation in BECs (see Supplementary Information for further discussion).

We simulate condensate formation using the stochastic Gross-Pitaevskii equation (SGPE) formalism \cite{Gardiner2002a,Gardiner2003a} that describes the highly-occupied, low-energy modes of a Bose gas with a classical field. The field evolves according to a generalised Gross-Pitaevskii equation that includes dissipation and thermal noise describing collisions between the partially condensed matter waves and the high-energy atoms in the thermal cloud.  Because evaporative cooling is difficult to simulate realistically \cite{Davis2002b}, and the details are often qualitatively unimportant, we implement an idealised cooling model with a sudden jump in chemical potential and temperature of the thermal cloud through the condensation critical point.  This leaves the SGPE classical field out of equilibrium with the thermal cloud; the subsequent return to equilibrium results in condensate formation.  Figure~\ref{fig1}a shows the growth in condensate number for the simulations of both quenches.  In our simulations, the initial and final thermal cloud parameters have been chosen to match the experimental results, and the coupling between the thermal cloud and the classical field is then adjusted to give good agreement with the experimentally observed BEC growth curves. This approach allows a meaningful comparison of other observables such as vortex statistics with the experimental data.  Further discussion can be found in Methods and Supplementary Information.

As shown in Figs.~\ref{fig2}b--c, vortices spontaneously form in our simulations, where each realisation can be interpreted as the numerical analog of a single experimental run.  We therefore study vortex dynamics in each growing condensate to compare vortex formation statistics with our experimental results.  In both our laboratory and numerical procedures, for each quench we repeat the BEC creation procedure and analyse statistics of vortex observations.  For each data set described below, we extract the fraction of images showing at least one vortex core within a displacement of $0.8 R_{TF}$ from the BEC centre, where $R_{TF}$ is the BEC Thomas-Fermi radius in the $z=0$ plane \cite{Pethick2002a}. This fraction serves as our estimate of the probability of observing spontaneously formed vortices in a single run.

Because localised decreases in the density profile of an experimentally obtained image may not always clearly indicate the presence of a core (e.g. due to tilting or bending with respect to the $z$ axis) our experimental uncertainty ranges are defined by our ability to visually determine whether or not an image shows a vortex.  For Quench A, 23\% to 28\% of 90 total images contain at least one visible vortex core.   For Quench B, 15\% to 20\% of 98 total images show at least one core.  Although the two quenches utilise quite different rf evaporation trajectories, they exhibit similar cooling and BEC growth rates. We can thus expect statistical similarities between the two data sets. Further statistical details, including results of observing multiple cores per image, are given in Supplementary Information.

From our simulations, we additionally analyse vortex observation probabilities as a function of time for each quench.  To determine the presence of a vortex we consider an instantaneous slice of the classical field in the $z=0$ plane of the trap, and detect all phase-loops of $\pm2\pi$ within a displacement of $0.8 R_{TF}$ from the BEC centre (here, $R_{TF}$ is based on the time-dependent condensate number).  We find that the majority of vortices are aligned with the $z$ axis of the trap.  The vortex observation probabilities obtained from $\sim$300 simulation runs for each quench are plotted against time in Fig.~\ref{fig1}b, with comparison to the experimental statistics.

According to our simulations the number of vortices decreases with time, consistent with our model of a thermal bath that is independent of time and that has no angular momentum; the thermodynamic final state should therefore be a condensate without vortices. In this respect the simulations diverge from our experimental observations, where there is no significant variation of vortex observation probabilities over a timescale of a few seconds. This low damping rate is consistent with the comparatively small thermal atomic component observed, indicating that a kinetic theory of thermal cloud dynamics may be needed to fully account for the long-time behaviour of the experiment.  We therefore compare our simulation results at $t=3.5$ s for Quench A, and $t=1.5$ s for Quench B, based upon experimental observations of negligible vortex damping once the BEC is nearly fully formed.

By focusing a blue-detuned laser beam propagating along the $z$ axis into the centre of the trap (see Methods), we experimentally studied BEC growth in a toroidal potential in which a BEC may display both persistent superfluid current \cite{ryu2007opf} about the central barrier, as well as free vortices circulating around the barrier.  The pinning of superfluid flow may influence both vortex dynamics during BEC growth and observations of vortices after the BEC is formed: a vortex pinned to the barrier reduces the likelyhood of complete self-annihilation between pairs of spontaneously formed vortices of opposite charge, thereby increasing the probability of finding a vortex in a BEC.  We implement a 6-s final evaporative cooling ramp identical to Quench A, and identify this data set as Quench C. An \emph{in-situ} image of a BEC in the toroidal trap is given in the leftmost image of Fig.~\ref{fig2}d.  Note that the dark region in the BEC centre indicates atoms displaced by the laser beam; vortices are not visible in this image.  After creating each BEC, we ramp down the laser power over 100 ms and immediately thereafter allow the BEC to expand from the trapping potential. For these conditions, we found 56\% to 62\% of 52 images contained at least one visible core; examples are shown in Fig.~\ref{fig2}d.

Condensate formation rates were not experimentally measured for Quench C; for the simulations we use the parameters of Quench A but with an additional repulsive Gaussian barrier with a height of 33 $\hbar \omega_r$. Simulated condensate growth versus time resulted in smaller condensates compared to Quench A (also observed experimentally) as shown as the green dot-dashed curve in Fig.~\ref{fig1}a.   Examples of numerically obtained column density and phase are shown in Figs.~\ref{fig2}e--f.  Vortex observation statistics for 300 runs are plotted as a green dot-dashed line in Fig.~\ref{fig1}b; we find that the vortex observation probability is approximately twice that of the harmonic trap of Quench A, as is also the case with the experimental data, but somewhat lower in overall magnitude than the experimental observations.  In contrast to the harmonic case, the curve does not exhibit decay below 40\% --- this corresponds to vortices pinned by the central barrier. Additional toroidal trap statistics are provided in Supplementary Information.

\begin{figure}
\includegraphics{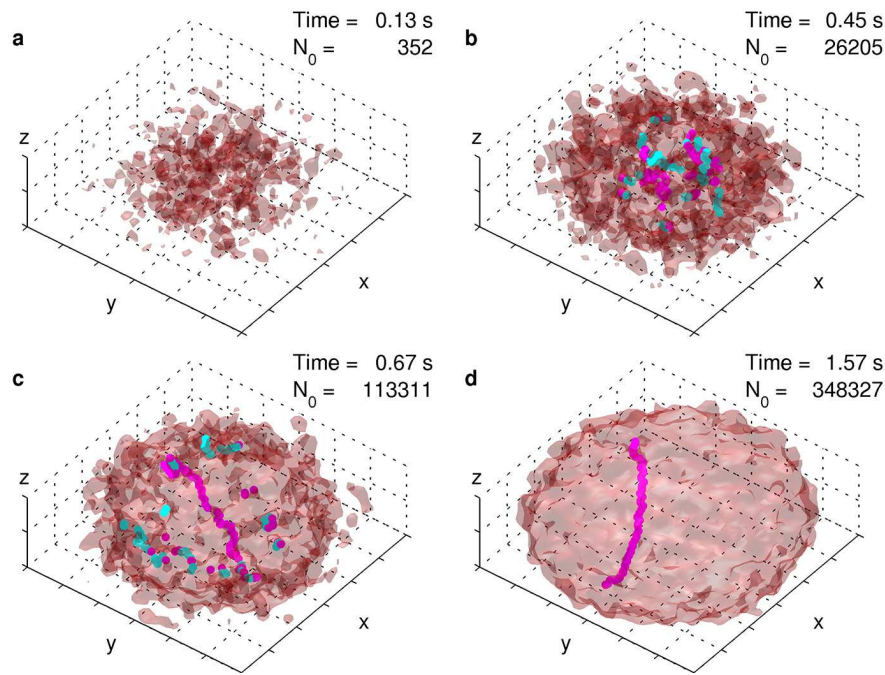}
\caption{\label{figmovieframes}\textbf{$|$~~BEC growth dynamics.}   \textbf{a--d}, Four snapshots during the simulated growth of a BEC showing isodensity surfaces (in light red) in a three-dimensional rendering.  Vortex cores of opposite charges about the $z$ axis are indicated  as magenta and cyan lines.  The corresponding times are \textbf{a}, 0.13 s;  \textbf{b}, 0.45 s;  \textbf{c}, 0.57 s;  \textbf{d}, 1.57 s, where $t=$0 s is the time when the quench is initiated in the simulation.   The full movie from which these images were taken is provided as Supplementary Video 3.
}
\end{figure}

In the Supplementary Information, we provide movies of simulated condensate formation for Quenches A and C. Here we describe one run in which a single vortex persists to the final time step for a Quench A simulation in the harmonic trap.   After the system temperature is initially lowered, the atomic field density profile fluctuates temporally and spatially, as illustrated in Fig.~\ref{figmovieframes}a. A bulk BEC then begins to grow, and a tangle of vortices is trapped within the BEC as shown in Fig.~\ref{figmovieframes}b, in qualitative agreement with the models of superfluid turbulence \cite{Svistunov1991,Kagan1992,Kagan1994,Kagan1997,Svistunov2001a,Berloff2002a} and the KZ mechanism \cite{Kibble1976a,Zurek1985a,Zurek1996a,Anglin1999a}.  At later stages, a nearly uniform condensate exists with clear vortex cores, as shown in Fig.~\ref{figmovieframes}c.  This state eventually damps to a single core, seen in Fig.~\ref{figmovieframes}d.

In examining the relationship between the BEC transition and the KZ mechanism, one would ideally study vortex formation with widely varying BEC growth rates in order to test the predictions of the scaling of the vortex density \cite{Kibble1976a,Zurek1985a,Zurek1996a}.  However, in our harmonic traps, we have only succeeded in increasing BEC growth rates by a factor of two to three, resulting in a factor of $\sim$2 increase in vortex formation compared with Quenches A and B.  In simulations, faster growth rates can potentially result in more vortices, as discussed in Supplementary Information.

In addition to providing new experimental observations, our work places spontaneous topological defect formation on a theoretical foundation that has not been available in analogous studies in other systems.  Related experimental investigations of spontaneous symmetry breaking of quenched ferromagnetic spinor BECs \cite{Sadler2006a} may also yield insight into phase transition dynamics.  The quantitative agreement between our experimental and theoretical results is of primary importance for their mutual interpretation: even in the ultra-cold BEC phase transition, thermal fluctuations can play an important role, and spontaneous topological defect formation may be virtually unavoidable in some situations.  Our continuing work will explore in greater detail exactly how a condensate forms in this regime; while the superfluid turbulence model describes vortex formation during condensation, to what extent does the KZ mechanism's universality relate to the BEC transition?  With further simulations and experiments, new understanding of the development of coherence in the birth of a superfluid may be uncovered, a tantalizing prospect addressing the interface between the classical and quantum worlds.

%%%%%%%%%%%%%%%%%%%%%%%%%%%%%%%%%%%%%%%%%%%%%%%%%%%%%%%%%%%%%%%%%%%%%%%%%%%%%
%%% METHODS SUMMARY
%%%%%%%%%%%%%%%%%%%%%%%%%%%%%%%%%%%%%%%%%%%%%%%%%%%%%%%%%%%%%%%%%%%%%%%%%%%%%
%
\section*{METHODS SUMMARY}

In our experiments, $^{87}$Rb atoms in the $|F=1, m_F=-1\rangle$ hyperfine state are confined in a time-averaged orbiting potential (TOP) magnetic trap \cite{petrich1995stc}.  Evaporative cooling increases the phase space density to near the condensation critical point. The trap frequencies are then relaxed, and a final stage of cooling (Quench A, B, or C) induces the phase transition.  In our numerical approach, the field consisting of the condensate and the low-energy, highly occupied modes of the gas is coupled to a bath of thermal atoms, parameterised with a chemical potential $\mu$ and temperature $T$ above a cut-off energy $E_{\rm cut}$.  Further details regarding our approaches are provided in the Methods section below.

%%%%%%%%%%%%%%%%%%%%%%%%%%%%%%%%%%%%%%%%%%%%%%%%%%%%%%%%%%%%%%%%%%%%%%%%%%%%%
%%% METHODS
%%%%%%%%%%%%%%%%%%%%%%%%%%%%%%%%%%%%%%%%%%%%%%%%%%%%%%%%%%%%%%%%%%%%%%%%%%%%%

\section*{METHODS}

\noindent\textbf{Evaporative cooling.} During the main evaporative cooling stages of our experimental procedure, our TOP trap is created with a spherical quadrupole field that has a vertical magnetic field gradient of $B_{z}' = 300$ G/cm, and a %$B_0=41$-G
magnetic bias field $B_0$ that has a direction that rotates in a horizontal plane at a frequency of either $\omega_{rot}=(2\pi)\cdot 4$ kHz or $\omega_{rot}=(2\pi)\cdot 2$ kHz.  Evaporative cooling %then
proceeds over 72 seconds as $B_0$ decreases from 41 G to $\sim$5.2 G, leaving a trapped cloud of atoms just above the condensation critical temperature $T_c$. $B_z'$ is then adiabatically reduced to $\sim$54 G/cm over 2 seconds, weakening the harmonic oscillator trapping frequencies to a measured radial (horizontal) trapping frequency of $\omega_r = 2\pi \cdot 7.8(1)$ Hz and an axial (vertical) trapping frequency of $\omega_z = 2\pi \cdot 15.3(2)$ Hz.  The center-of-mass position of the atom cloud correspondingly sags by approximately 0.6 mm vertically.  In the final stage of our cooling cycle for Quench A, we use a continuous 6-s ramp of the rf field, which induces the evaporative cooling of the atom cloud from 70 nK to 20 nK, with $T_c \sim 42$ nK, to create condensates of $N_c \sim 5\times 10^5$ atoms. For Quench B, the continuous rf evaporative cooling ramp is replaced with a sudden rf jump to a final rf value, followed by a hold of the atomic sample in the trap before release and imaging.  In this situation we find $T_c \sim 35$ nK and the final condensate number is $N_c \sim 3\times 10^5$ atoms.

\noindent\textbf{TOP trap.}  To ensure that the rotating bias field of the TOP trap plays no significant role in the spontaneous formation of vortices, we measured the $z$-component of the net orbital angular momentum $L_z$ of our condensates using surface wave spectroscopy.  We excite a quadrupolar oscillation of the BEC in the horizontal plane, and stroboscopically probe the BEC with a set of non-destructive in-trap phase-contrast images, obtained by probing along the $z$ axis \cite{Chevy2000mam,Haljan2001usw}.  The quadrupolar oscillations will then precess with a rate and direction proportional to $L_z$.  In our measurements, there was no significant biasing of surface mode precession in a direction corresponding to the TOP trap rotation direction, an indication that TOP trap temporal dynamics have little to no influence on spontaneous vortex formation.  This is discussed further in Supplementary Information.

\noindent\textbf{Toroidal trap.}  A potential-energy barrier was added to the centre of the magnetic trap using a focussed blue-detuned laser beam with a wavelength of 660 nm, $\sim$18 $\mu$W of power, and a 6-$\mu$m Gaussian radius. The maximum beam intensity corresponds to a potential energy of approximately k$_B\cdot$20 nK, where k$_B$ is Boltzmann's constant.  This can be compared with a $\sim$k$_B\cdot$10 nK chemical potential of our fully formed BECs in the purely harmonic trap.  The beam was adiabatically ramped on prior to the final 6-s evaporation ramp, only slightly perturbing the thermal cloud but providing enough additional potential energy along the trap axis to exclude BEC atoms from the $z$ axis of the trap.

\noindent\textbf{Imaging.}  Our main imaging procedure involves the sudden removal of the magnetic trap, and the subsequent ballistic expansion of the trapped cloud.  After 59 ms of expansion in the presence of an additional magnetic field to support the atoms against gravity, the atomic cloud is illuminated with near-resonant laser light, and the absorption profile of the atomic density distribution is imaged onto a camera.  In our grey scale images, lighter shades represent higher optical depth, proportional to integrated column density along the $z$-direction line-of-sight.   A clear vortex core aligned along the $z$ axis appears as a dark hole in the density distribution.

\noindent\textbf{Stochastic Gross-Pitaevskii theory.}  We denote the condensate and low-energy portion of the trapped gas with the field  $\alpha(\mathbf{x},t)$, and define the Gross-Pitaevskii operator
\begin{eqnarray}
L_{\rm GP} = -\frac{\hbar^2}{2m}\nabla^2 + V(\mathbf{x}) +
g |\alpha(\mathbf{x},t)|^2,
\end{eqnarray}
where $m$ is the mass of an atom, $V(\mathbf{x})$ is the trapping potential, $g = 4\pi
\hbar^2 a/m$ characterises the strength of atomic interactions, and $a$ is the s-wave scattering
length. The equation of motion for the field is
\begin{eqnarray}
d\alpha(\mathbf{x},t) &=&{\cal P}\left\{-\frac{i}{\hbar}  L_{\rm GP}\alpha(\mathbf{x},t) dt +\frac{G(\mathbf{x})}{k_BT}(\mu-L_{\rm GP})\alpha(\mathbf{x},t) dt+dW_G(\mathbf{x},t)\right\},\nonumber
\label{SGPEaloc}
\label{eq:SGPE}
\end{eqnarray}
which has been derived from first principles using the Wigner phase-space representation \cite{Gardiner2003a}.
The first term on the right describes unitary evolution of the classical field according to the Gross-Pitaevskii equation. The second term represents growth processes, i.e.\ collisions that transfer atoms from the thermal bath to the classical field and vice-versa, and the form of $G(\mathbf{x})$ may be determined from kinetic theory \cite{Bradley2007a}. The third term is the complex-valued noise associated with condensate growth. The noise has Gaussian statistics and is defined by its only non-vanishing moment:
$\langle dW_G^*(\mathbf{x},t) dW_G(\mathbf{x}',t') \rangle =
2 G(\mathbf{x})dt \;\delta(\mathbf{x} - \mathbf{x}')\delta(t-t')$; it is also consistent with the fluctuation-dissipation theorem.
The projection operator ${\cal P}$ restricts the dynamics to the low-energy region \cite{Davis2001b,Blakie2005a} defined by all harmonic oscillator modes with energy $\epsilon < E_{\rm cut} = 40 \hbar \omega_r$ for these calculations, which for our parameters gives about three particles per mode at the cutoff.  For typical experimental parameters this method is accurate from slightly above the critical temperature to colder temperatures where there is still a significant thermal fraction \cite{Davis2006a}.

The initial states used in our simulations are independent field configurations generated by ergodic evolution of the SGPE at equilibrium with the thermal cloud
with $\mu_i = \hbar \omega_r$ and $T_i = 45 (35)$ nK for Quench A (B), representing the thermalised Bose gas above the transition temperature \cite{Davis2006a}.  These parameters are then suddenly changed to final values chosen to match the final condensate number and temperature observed in the experiment: $\mu_f = 25$ $(22)$ $\hbar \omega_r$ and $T_f$ = 34 (25) nK for Quench A (B). We perform simulations for 300 (298) sets of initial conditions.  By averaging over the different realisations we can calculate any quantum-mechanical observable as a function of time, and in particular we diagonalise the single-particle density matrix to find the number of atoms in the condensate \cite{Blakie2005a}.%,Goral2002a}.

Because vortex formation is expected to depend upon the BEC growth rate, which is difficult to calculate precisely, we adjust the coupling rate describing Bose-enhanced collisions between the classical field and thermal cloud to obtain a close match of the experimental BEC growth curves.
We choose a spatially constant rate for the dimensionless coupling $\gamma = \hbar G(\mathbf{x})/k_BT$. The noise at each time step then has the explicit form $dW_G(\mathbf{x},t)=\sum \phi_j(\mathbf{x})\sqrt{\gamma k_BT dt/\hbar}(\eta_j+i\zeta_j)$, where $\phi_j(\mathbf{x})$ are the single particle modes below the cutoff and the real Gaussian variables $\eta_j$, $\zeta_j$, are  independent and have zero mean and unit variance. In principle $\gamma$ is specified by a quantum Boltzmann integral, but here we treat it as an experimental fitting parameter for the condensate growth rate; it is never more than a factor of two different from the result of Eq.~(A11) in Bradley \emph{et al}. \cite{Bradley2007a}. The effect of these parameter choices is discussed further in Supplementary Information.

%%%%%%%%%%%%%%%%%%%%%%%%%%%%%%%%%%%%%%%%%%%%%%%%%%%%%%%%%%%%%%%%%%%%%%%%%%%%%
%%% END OF MAIN TEXT
%%%%%%%%%%%%%%%%%%%%%%%%%%%%%%%%%%%%%%%%%%%%%%%%%%%%%%%%%%%%%%%%%%%%%%%%%%%%%

\vspace{1cm}
\noindent\textbf{References} section follows the Supplementary Information.\\

\subsection{Acknowledgements} We thank David Roberts, Boris Svistunov, Ewan Wright, and Wojciech Zurek for useful discussions.  The experimental work was supported by the US National Science Foundation under grant 0354977, and by the Army Research Office.  The theoretical work was supported by the Australian Research Council and the University of Queensland.

\subsection{Author Information}
% \item[Competing Interests]
The authors declare that they have no
competing financial interests.
% \item[Correspondence]
Correspondence and requests for materials
should be addressed to B.P.A. (bpa@optics.arizona.edu).
%\end{addendum}

%%%%%%%%%%%%%%%%%%%%%%%%%%%%%%%%%%%%%%%%%%%%%%%%%%%%%%%%%%%%%%%%%%%%%%%%%%%%%
%%% FIGURES
%%%%%%%%%%%%%%%%%%%%%%%%%%%%%%%%%%%%%%%%%%%%%%%%%%%%%%%%%%%%%%%%%%%%%%%%%%%%%

\newpage
\def\figurename{SUPPLEMENTARY FIG.}
\addtocounter{figure}{-4}

\begin{center}
\begin{large}
\textbf{Supplementary Information:\\
Spontaneous vortices in the formation of Bose-Einstein condensates}
\end{large}
\end{center}
%\vspace{5mm}
\begin{center}
C.~N.~Weiler$^1$, T.~W.~Neely$^1$, D.~R.~Scherer$^1$,
A.~S.~Bradley$^2$, M.~J.~Davis$^2$, \& B.~P.~Anderson$^1$\\

\begin{small}\textsl{$^1$College of Optical Sciences, University of Arizona,
Tucson, AZ 85721, USA\\
$^2$ARC Centre of Excellence for Quantum-Atom Optics,
School of Physical Sciences, University of Queensland,\\
Brisbane, Queensland 4072, Australia}
 \end{small}
\end{center}

\noindent\textbf{\large Supplementary Video Legends}\\

Here we describe the main features of a selection of movies of our condensate formation simulations.  Movie files are available online at \textbf{http://www.physics.uq.edu.au/people/mdavis/spontaneous\_vortices}. Ten example movies are given: five corresponding to condensate formation in the purely harmonic trap (Quench A), and five in the toroidal trap (Quench C).  As specified in the title frames, all simulations in this section have initial chemical potential $\mu_i = \hbar \omega_r$ and initial temperature $T_i = 45$ nK, and corresponding final values $\mu_f = 25 \hbar \omega_r$ and $T_f = 34$ nK.  Quench C has an additional central repulsive gaussian barrier of height $33 \hbar \omega_r$.

All movies begin at time $t=0$ corresponding to the instant that the chemical
potential and temperature are suddenly changed in order to induce condensation.
The left panel of each movie shows a rendering of a three-dimensional isosurface
of the density of the classical field, with magenta (cyan) dots indicating the
presence of positive (negative) phase windings about the $z$ direction.  We
indicate only the phase windings located in a spatial region where there is a
significant condensate density, as determined by the Penrose-Onsager criterion.
At earlier times, these points of fluid circulation seem to be distributed
almost at random and at high spatial density, however at longer times these
points can often be seen to form vortex lines that extend through the
condensate.  The top-right panel of each frame plots the column density along
the $z$-axis of the classical field.  The greyscale colourmap is fixed in time as indicated by the gradient bar at the far right (arbitrary units), and the numerical data appear similar to the type of images obtained in the experiment, although without the time-of-flight expansion.  Vortices can often be seen as holes or low density regions in the column density image depending on their orientation with respect to the $z$-axis.  The lower right panel shows the phase of the field in the $z=0$ plane.  Positive (negative) phase windings are indicated as magenta crosses (cyan circles) in the regions of significant condensate density.  It should be noted that many additional $2\pi$ windings in this panel occur where there is no condensate density, and are hence not to be interpreted as vortices.\\[5mm]
\noindent\textbf{Condensate formation movies in a harmonic trap: Quench A.} From 300 simulations of Quench A, 229 contained no vortex cores after 1.5 s. The following five movies show a selection of outcomes, including cases with 0, 1, and 2 long-lived vortices.\\

\noindent\textbf{Supplementary Video 1:}
 The first movie is an example of BEC formation in which there are no vortices trapped in the final condensate, and is provided for comparison purposes.  At early times there is some indication of vortices in the condensate; however, these do not survive at longer times.  In this situation the condensate appears to grow adiabatically in its ground state, as was assumed in the condensate formation calculations of Refs.~\citenum{QKPRLI,QKPRLII,Lee2000a,Davis2000a,Bijlsma2000a}.
The reader should note that cases such as this are typical for the data set corresponding to Quench A.\\

\noindent\textbf{Supplementary Video 2:}
This movie provides an example of BEC formation with a single vortex line that survives at long times and remains close to the centre of the condensate.  The vortex line remains approximately vertical, and is usually easily visible in the column density plot.  The reader should note that there are many examples similar to this where the vortex line is not so close to the centre, and instead slowly spirals towards the boundary of the condensate over a time scale of several seconds.\\

\noindent\textbf{Supplementary Video 3:}
In this example there are three clear vortices at early times.  Two of these arise near the edge of the condensate and these damp out relatively quickly, leaving a single vortex line of opposite charge near the centre that survives to the end of the simulation.  This is an example of a vortex that does not directly align with the $z$ axis, and the column density often shows an elongated density dip.  This movie uses the same simulation data as in Figure 5 of the main text.\\

\noindent\textbf{Supplementary Video 4:}
This simulation results in two oppositely charged vortices that remain well separated, precessing about the centre in opposite directions.  It should be noted that only 3 out of 300 simulations in this data set clearly exhibited more than one core at long times.\\

\noindent\textbf{Supplementary Video 5:}
The final movie for the data set of Quench A shows a number of vortex cores that undergo some rather complicated dynamics as they move about within the condensate, including vortices that cross each other and reconnect in a different configuration, and an example where a vortex line flips (and hence changes colour).  This example shows the most complicated vortex dynamics observed in this data set.\\[5mm]

\noindent\textbf{Condensate formation movies in a toroidal trap: Quench C.}
From the 300 simulations of Quench C, 147 contained no vortices after 1.5 s.  The following movies show five examples where vortices arose.\\

\noindent\textbf{Supplementary Video 6:}
The first movie from this data set shows the formation of a single vortex.  This is not readily seen in the isodensity surface, as the $2\pi$ phase winding is located in the centre of the trap where there is no condensate density.  However, the phase winding about the entire condensate is clear in the bottom-right plot of the phase in the $z=0$ plane.
Whereas in the harmonic trap all vortices will eventually make their way to the edge of the condensate and disappear, here the gaussian barrier pins the vortex to the centre --- this is an example of a persistent current.\\

\noindent\textbf{Supplementary Video 7:}
In this example it is possible to see a pair of oppositely charged vortices at $t=0.8$ s, one of which becomes pinned to the central barrier while the other ends up precessing about the centre in the region of maximum condensate density.\\

\noindent\textbf{Supplementary Video 8:}
Here there are some rather complicated early dynamics that result in a single vortex precessing about the outside of the condensate, but with no persistent current.\\

\noindent\textbf{Supplementary Video 9:}
There are a number of vortices early on, and near 0.6 s it seems that there is a doubly charged persistent current.  However, only one vortex ends up being trapped on the central gaussian barrier and the other vortex of the same charge remains within the bulk of the condensate, precessing rapidly about the centre.  The rate of precession should be compared with the example from Supplementary Video 7, where the vortex near the edge is of the opposite charge to that pinned by the barrier.\\

\noindent\textbf{Supplementary Video 10:}
The final example shows a stable doubly-charged persistent current.  This is particularly interesting given the energy difference between this and the thermodynamic ground state of the condensate with no current.  This is the only example from the data set where a stable doubly-charged persistent current was seen.\\[5mm]

\noindent\textbf{\large Supplementary Discussion}\\

\noindent\textbf{Vortex distribution in the toroidal trap.}
We observe in the simulated BEC growth dynamics for the toroidal trap that the central potential can pull in and pin a single spontaneously formed vortex in some cases.  By examining the distribution of vortex core positions for both the experimental and simulated data we see that the toroidal trap induces a central clustering of vortex cores, whereas in the harmonic trap the core positions are more evenly spread throughout the BECs.  A visual comparison of these cases is given in Supplementary Figs.~\ref{fig4}a--d. Supplementary Figures~\ref{fig4}e--f show a histogram of vortex core displacements away from the mean core position for the two different trap geometries for theory and experiment.  It should be noted that the experimental images in Supplementary Figs.~\ref{fig4} are taken after the removal of the central barrier of the toroidal trap (if present), and after time-of-flight imaging, whereas the simulated images correspond to in-trap results.  Hence, the data in this figure should be taken as indicative only of trends between harmonic versus toroidal trap results.  Despite this, the vortex position data show some agreement between the simulations and the experimental data.  Due to the observation that the cores in the experimental toroidal trap data are likely to be found within a small region within the BEC, we interpret our results as indicating that the observed cores are likely to have been pinned to the central barrier prior to expansion.  This suggests possibilities for future controlled studies of spontaneous vortex formation, perhaps with multiple sites at which vortices may be pinned, in order to better understand the density and numbers of vortices created at early times in BEC growth.
\begin{figure}
\includegraphics[width=105mm]{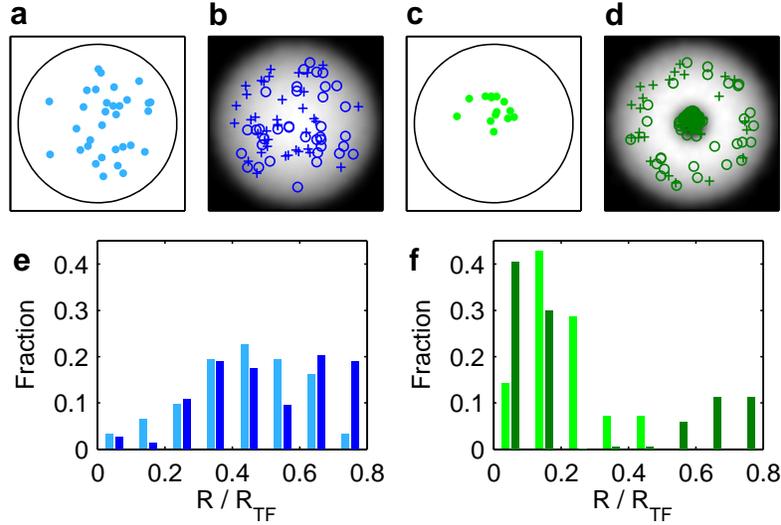}
\caption{\label{fig4}\textbf{Vortex core pinning.} \textbf{a, c}, Representation of the experimentally measured positions of the vortex cores relative to the Thomas-Fermi radius $R_{TF}$ (outer circles) for \textbf{a}, the harmonic trap and \textbf{c}, the toroidal trap. \textbf{b, d}, Corresponding theoretical results, crosses and circles indicate oppositely charged vortices. \textbf{e, f}, Comparison of the statistics of the vortex displacements $R$, binned in steps of $R / R_{TF} = .01$, for the experimental data (left bar in each pair) and theoretical simulations (right bars).  Harmonic trap results are shown in \textbf{e}, toroidal trap results are shown in \textbf{f}.  For the experimental data, only images clearly showing a single core within 0.8 of the Thomas-Fermi radius were considered.  For the data represented in this figure, the experimental data are obtained from \emph{expanded} images, and the numerical data are obtained from \emph{in-trap} simulations.  In \textbf{d}, the individual vortex cores pinned to the central barrier can not be separately resolved; nevertheless, the total number of cores pinned to the barrier significantly outweigh the number that are not pinned.  Thus for both experimental and simulated toroidal, this figure shows that the toroidal trap strongly influences the vortex core positions within the BEC, and that cores are most likely to be found within a smaller region of the BEC for the toroidal trap than for the harmonic trap.
}
\end{figure}

We also note that the presence of a repulsive light field to create a toroidal potential could in principle lead to mechanisms for vortex formation other than spontaneous vortex formation during the BEC phase transition, particularly if mechanical stirring exists (for instance, if the beam or harmonic potential is not stationary).  We do not believe that a stirring mechanism is at work in our experiment.  Through phase-contrast images, we confirmed that residual center-of-mass motion of the BEC (once formed) is minimal, with BEC position varying by approximately $\pm$2 $\mu$m at most, and usually much less.  This may be due to our procedure of evaporatively cooling directly in a final weak trap, rather than making a BEC in a tighter trap and then relaxing to a weak trap (larger residual center-of-mass oscillations may be observed when relaxing the trap after making a condensate).  Furthermore, our laser beam position is relatively stable with respect to the center of the harmonic potential throughout the course of a full data set, and fluctuates or drifts less than 4 $\mu$m during an experimental run.  We therefore believe that unintended stirring of the BEC by the laser beam is unlikely.\\[5mm]

\noindent\textbf{Vortex statistics.}  The table below shows a numerical comparison between the experimental and numerical vortex observation statistics in greater detail than provided in the main text.  For each quench condition with total number of runs listed, $n_j$ shows the number of times $j$ vortices were observed, and $P_j$ shows the corresponding probability of observing $j$ vortices in a single run.  In the experimental toroidal trap data, 1--2 images, or 2\%--4\%, showed 3 cores.  This is not represented in the table.\\
\begin{center}
\begin{tabular}{|c|c|c|c|c|c|c|c|} \hline
\textbf{Quench} & \textbf{runs} & \multicolumn{2}{c|}{\textbf{0 cores}} & \multicolumn{2}{c|}{\textbf{1 core}} & \multicolumn{2}{c|}{\textbf{2 cores}} \\ \cline{3-8}
& & \textbf{n}$_0$ & \textbf{P}$_0$ & \textbf{n}$_1$ & \textbf{P}$_1$ & \textbf{n}$_2$ & \textbf{P}$_2$ \\ \hline\hline
A, expt. & 90 & 65 -- 69 & 0.72 -- 0.77 & 18 -- 23 & 0.20 -- 0.26 & 2 - 3 & 0.02 -- 0.03 \\ \hline
A, sim. & 300 & 229  & 0.76 & 68 & 0.23 & 3 & 0.01 \\ \hline\hline
B, expt. & 98 & 78 -- 83 & 0.80 -- 0.85 & 13 -- 18 & 0.13 -- 0.18 & 2 & 0.02 \\ \hline
B, sim. & 298 & 234  & 0.79 & 61 & 0.20 & 3 & 0.01 \\ \hline\hline
Toroidal, expt. & 52 & 20 -- 23 & 0.38 -- 0.44 & 15 -- 25 & 0.29 -- 0.48 & 5 -- 12 & 0.10 -- 0.23 \\ \hline
Toroidal, sim. & 300 & 147  & 0.49 & 137 & 0.46 & 16 & 0.05 \\ \hline
\end{tabular}
\end{center}
\begin{center}
\textbf{Supplementary Table 1:~~Harmonic and toroidal trap vortex observation statistics.}\\
\end{center}

From Fig.~2b in the main text and from Supplementary Table 1 it can be seen that there is some difference between the experimentally estimated probability of vortex observation compared to that from the simulations.  In particular the simulation estimate is slightly higher than experiment for the sudden cooling of Quench B, and slightly lower than experiment for the 6-s evaporative cooling of Quenches A and C at the time of observation.  Due to the limited number of realisations of the experiment for these three runs, it is difficult to say whether this is real difference or due to statistical fluctuations.  The theoretical modelling of the evaporative cooling is quite basic, and there is a strong possibility that this could affect the theoretical results in some manner.  However, further simulations (discussed below) suggest that the simulations results are quite robust to the time scale of the quench of $\mu$ and $T$.

Finally, we again note that the condensate growth curve was not measured experimentally for Quench C, which shows the largest discrepancy in vortex statistics between theory and experiment.  It is possible that our simulations may not accurately represent the rate of growth of the condensate or the final condensate number for this case, giving a further plausible explanation for the difference.

\vspace{5mm}
\noindent\textbf{Vortices in phase transitions in ultra-cold gases.}  The description of defect formation in the Kibble-Zurek scenario is appealing due to its simplicity and universality.  However, a somewhat more complete description for the quantum gas is the evolution from weak turbulence to a quasi-condensate and
superfluid turbulence as described by Svistunov and co-workers [9,18-21].
(for a summary see Ref.~\citenum{Svistunov2000}).
In this scenario, as the system cools (but before condensation), the low-energy modes of the gas become highly occupied, and can be treated as classical waves.  The superposition of a number of low-energy modes with large amplitudes and random phases (as appropriate for a thermal distribution) naturally gives zeros in the density of the gas via destructive interference, which can be considered as precursors to vorticity.  A rapid quench leading to condensate formation can freeze these defects into the resulting superfluid, giving a tangle of vortices that then relaxes over a relatively long time scale.  During this period of time, the system is not in thermodynamic equilibrium.  The presence of vortices suppresses the establishment of long-range order, and the system is said to be in a \textit{quasi-condensate} state. However, the thermodynamic state of the system that would be realised if the cooling proceeded more slowly is a condensate without vortices --- it is the rapid dynamics of condensate growth that can trap defects in the superfluid.

Vortices are present in a two-dimensional (2D) Bose gas at cold temperatures (but above any superfluid critical temperature) for a similar reason --- there is a superposition of highly occupied classical modes with random phases.  This is why classical field models (related to our theoretical techniques) have had success in modelling Berezinskii-Kosterlitz-Thouless (BKT) experiments in Bose gases \cite{Hadzibabic2006a,Schweikhard2007a},
see for example Refs.~\citenum{Simula2006a,Simula2008a,Bisset2008a}.
However, in a homogeneous system at least, the thermodynamic state of the system below the BKT superfluid temperature contains vortices that arise in bound pairs.  The superfluid-to-normal fluid transition occurs when the temperature is sufficiently high that these pairs start to unbind, resulting in free vortices.   The case of a trapped gas seems to be more complicated than this, and is only now beginning to be understood; see for example Bisset \emph{et al.} \cite{Bisset2008a} for a recent discussion of vortices, superfluidity, condensation and long-range order in 2D trapped systems.  Because vortices and vortex dynamics are an important element of the BKT transition, our work on spontaneous vortex formation may at first appear somehow related to the BKT transition.  However, the main difference between the BKT transition and spontaneous vortex formation can be briefly summarized as follows.  Both above and below the critical temperature of the BKT transition, the thermodynamic state of a cold homogeneous 2D gas contains vortices (either free or bound in pairs); the BKT transition involves a transition from one equilibrium state (with paired vortices) to another equilibrium state (with free and unbound vortices).  In our work involving spontaneous vortex formation in a three-dimensional gas, vortices are seen because they are frozen into the evolving gas (the quasi-condensate) during a fast transition when the system is out of equilibrium.  The thermodynamic equilibrium state below the critical point is free of vortices, and vortices are only observed in our work due to rapid-enough cooling in which the system falls out of equilibrium, and by observing the system during its relaxation from a quasi-condensate to a true condensate.\\[5mm]

\noindent\textbf{Estimating the correlation length.} Here we provide a brief summary of our procedure to estimate the correlation length $\xi$ relevant for the BEC phase transition of our experiments.  This method follows the procedure outlined by Anglin and Zurek [8].
In the KZ model, a system undergoing a phase transition will fall out of thermal equilibrium when the equilibration (or relaxation) timescale $\tau$ exceeds the time remaining before the critical point is reached.  This essentially means that the decreasing equilibration rates can not keep up with the cooling rate near the phase transition critical point.  Defining time $t=0$ as the time when the system reaches the critical point, the time $t = -\hat{t}$ is defined as the point at which the system falls out of equilibrium.  The time $\hat{t}$ can be estimated through the relationship $\hat{t} = \sqrt{\tau_Q\tau_0}$, where $\tau_Q$ is the quench time ($1/\tau_Q$ is the quench rate as defined in our paper), and $\tau_0$ is the scattering time of atoms in the gas.  For our data, $\tau_0 \sim 0.1$ s, and $\tau_Q \sim 7$ s for Quench A and $\sim$5 s for Quench B, where $\tau_Q$ is obtained from a linear fit of the function $(1 - t/\tau_Q)$ to the temperature data of Figure 2a (inset) of the main text.  This leads us to an estimate of $\hat{t} \sim 0.8$ s.  In other words, the correlation length $\xi$ of the gas at $t = -0.8$ s becomes essentially frozen into the system because the gas can not regain equilibrium until a time $t = 0.8$ s \emph{after} passing through the critical point.  We estimate $\xi$ via the relation $\xi \approx \lambda_{\mathrm{dB}} (\tau_Q/\tau_0)^{1/4}$, where $\lambda_{\mathrm{dB}}$ is the thermal deBroglie wavelength of the system at $t = -\hat{t}$.  For our parameters, $\xi \sim 0.6$ $\mu$m for Quenches A and B.\\[5mm]

\noindent\textbf{Experimental vortex observations.}
In our experiments on condensate formation we have not performed any uncommon experimental techniques in order to observe spontaneously formed vortices. The Bose gas of $^{87}$Rb atoms was evaporatively cooled in a similar manner to those of most other dilute-gas BEC experiments [27].
One might assume that rapid cooling is necessary in order for distinct regions of coherence and hence spontaneous vortices to form, in contradiction with the seemingly long cooling time scales for Quenches A and C. It is then natural to wonder why our experiments have yielded observations of spontaneous vortex formation during the BEC phase transition given that this has not been reported previously.  We believe that the major factors contributing to our observations are as follows:
\begin{enumerate}
\item We evaporatively cool our gas to near-degeneracy in a tight trap, and then relax the trap to the final trapping frequencies before our final cooling ramp.  This relaxation reduces the atomic collision rate at the point at which condensation occurs, and potentially decreases the ratio of the correlation length to the size of the ground state of the trap (the harmonic oscillator length) and thus increases the number of vortices expected to form.  In the KZ mechanism, the correlation length $\xi$ depends on the quench time $\tau_Q$, so specific details of the ratio of correlation length to the harmonic oscillator length will depend upon specific cooling trajectories, as well as trap frequencies.  Regardless of the initial number of vortices formed, we expect that vortices may survive for longer times in weaker traps, where the lower atomic density and longer trap oscillation time scale generally decrease the rates of dynamical processes.
\item Our experiment is performed in an oblate trap with an approximately 2:1 aspect ratio, in which it is favourable for the vortices to align along the vertical symmetry axis of the trap \cite{svidzinsky2000dvt}. All analogous condensate formation dynamics experiments reported to-date in the literature have been performed in cigar-shaped traps \cite{Shvarchuck2002a,Miesner1998a,QKPRLIII,Hugbart2006a,Ritter2006a}.
\item We introduce a linear magnetic field gradient during expansion that supports the condensate against gravity and allows for long expansion times before imaging.  During ballistic expansion of the condensate, the vortex cores expand relative to the condensate size such that our imaging system can optically resolve the cores.  Unless one is actively looking for vortex cores, such long expansion times (59 ms in our case) are not necessarily needed or desired in other experiments.
\item We have the ability to acquire images of trapped atoms along both the vertical axis (the direction of strongest confinement) as well as the horizontal axis, the more standard imaging axis for TOP trap experiments.  The vertical imaging axis allows us to detect vortices that are well-aligned with the vertical trap axis.
\end{enumerate}
It seems entirely possible that spontaneous vortices have been present in other experiments, but simply could not be resolved, damped quickly, or were not sought.\\[3mm]

\noindent\textbf{Influence of the TOP field.}
Using the method of quadrupolar surface mode excitation and precession detection [31,32],
early experimental results from this work appeared to show a significant bias of the direction of fluid circulation for the observed spontaneously formed vortices. Our initial measurements seemed to indicate that the direction of vortex fluid flow for the spontaneously formed vortices was most often found to be circulating in the same azimuthal direction as the rotating magnetic field that forms our TOP trap.  This suggested that some form of stirring or other influence from the TOP trap was responsible for the apparent directional bias.  However, our later and more detailed measurements on condensates created with Quench A revealed that the apparent bias was a result of a scissors mode oscillation that was unintentionally excited.  Coupled with the times that we initially chose to stroboscopically probe the surface mode precession, the scissors mode oscillation appeared as a bias to the precession direction.  Such directional biasing is not detected in our more recent measurements using Quench A, and we do not believe that the rotating field of the TOP trap plays a role in vortex formation (these measurements were not performed with the toroidal trap data). Furthermore, there is no observable difference in vortex observation statistics between data obtained with a 2 kHz or with a 4 kHz rotating bias field for the TOP trap; if the field were somehow stirring the cloud, it is likely that a difference in vortex observation statistics for the two cases should be detectable due to different angular momentum coupling between the rotating magnetic field and the trapped atoms.
\\[3mm]

\noindent\textbf{Relation to simulations of Berloff and Svistunov.}
Here we briefly compare the simulations of Berloff and Svistunov [22] to our own numerical calculations.  The fact that our simulations are for the trapped gas rather than the homogeneous gas means that there are many quantitative differences compared to those of Berloff and Svistunov [22] --- however the basic picture of strongly nonequilbrium dynamics resulting in a vortex tangle and superfluid turbulence before subsequent relaxation is the same.

Berloff and Svistunov [22] simulated the transition from weak to superfluid turbulence in a homogenous Bose gas using a discretised Gross-Pitaevsii equation on a large cubic grid of dimension $256^3$. Their initial condition was a superposition of plane waves according to
$$
\psi(\mathbf{r},t=0) = \sum_{\mathbf{k}} a_\mathbf{k} \exp(i\mathbf{k}\cdot\mathbf{r}),
$$
where the phases of the complex amplitudes $\{a_\mathbf{k}\}$ were randomly distributed as appropriate for the weakly-interacting Bose gas in the kinetic regime [21].  The subsequent dynamics resulted in the redistribution of the mode occupation numbers leading to a growth in the low-momentum modes and the formation of a quasi-condensate while conserving energy and particle number.  The lines of zero density present in the initial state evolved into a vortex tangle as the quasi-condensate formed, which subsequently relaxed further and resulted in a number of vortex lines in the final state.

In the visualisation of the resulting vortex networks, Berloff and Svistunov either filtered out the high momentum modes, or performed short-time averaging of the density,  resulting in a smoothing of the isodensity surfaces that were plotted.  In some sense the high-momentum modes provide a mechanism for dissipation as the low-momentum modes of the system evolved towards thermal equilibrium.

The main differences between the simulations of Berloff and Svistunov [22] and those performed in this paper are as follows:
\begin{enumerate}
\item Our simulations begin with an equilibrium state for the low-energy classical modes above the critical temperature for condensation.  The quenching of our thermal reservoir to a lower temperature and higher chemical potential results in the classical region evolving to a strongly non-equilibrium state similar to the initial condition of Berloff and Svistunov [22] appropriate to a trapped gas.  This subsequently evolves through the quasi-condensate and superfluid turbulent stages towards equilibrium, sometimes trapping vortices.  The thermal reservoir provides the mechanism for dissipation in our simulations.
\item We explicitly separate the high-energy modes from the classical field represented by the SGPE, and represent them as a thermal reservoir.  Berloff and Svistunov include a large number of what we would classify as high-energy modes in their microcanonical GPE approach.  If the occupation number of these modes is not sufficiently high, this is not an entirely quantitatively accurate representation of the full system.  However, the dynamics of the high-energy modes has little qualitative impact on the dynamics of the low-momentum modes that form the superfluid, and provide an effective dissipative mechanism for relaxation.
\end{enumerate}
In our opinion, the key innovations differentiating our simulations from those of Berloff and Svistunov [22] are the efficient implementation of harmonic oscillator basis in our representation of the classical field, and the inclusion of the high-energy modes using grand canonical classical field theory [11,12,14,15].
\\[5mm]

\noindent\textbf{Dependence of results on simulation fitting parameters.}
In the main text we have compared our experimental results to simulations in which the chemical potential $\mu$ and temperature $T$ underwent an instantaneous quench to their final values.  In this section we discuss the dependence of the simulations modelling Quench A on the parameters that characterise our idealisation of the cooling process.

The final temperature chosen for the simulations was based on the approximate value measured experimentally once the slope of the condensate growth curve began to flatten out.  We have also performed a set of 100 simulations where the final temperature was $T_f = 23$ nK instead of $34$ nK, and found no difference in the vortex probabilities within statistical noise.  This implies that for our simulations the time-scale of the temperature decrease and the exact final value has little effect.

We have also performed the same simulations for 100 trajectories using a linear ramp of the temperature and chemical potential from their initial to final values over time scales of $1$, $2$, $4$, $8$, $16$, and $24$ radial trap periods, and the results are shown in Supplementary Figures~\ref{SIfig2}a--b.  We find that the first four of these give condensate growth curves that cannot be distinguished, apart from an increasing delay in the initiation of growth.  The vortex observation probability appears to drop slightly with increasing ramp time, but this is difficult to distinguish within the statistical error.  As we are not actually modeling the evaporative cooling trajectory of the experiment, and the  time axis of our simulations is shifted to best fit the experimental data, it seems that the time scale of the quench is of little consequence for our simulations.

\begin{figure}
\includegraphics[width=78mm]{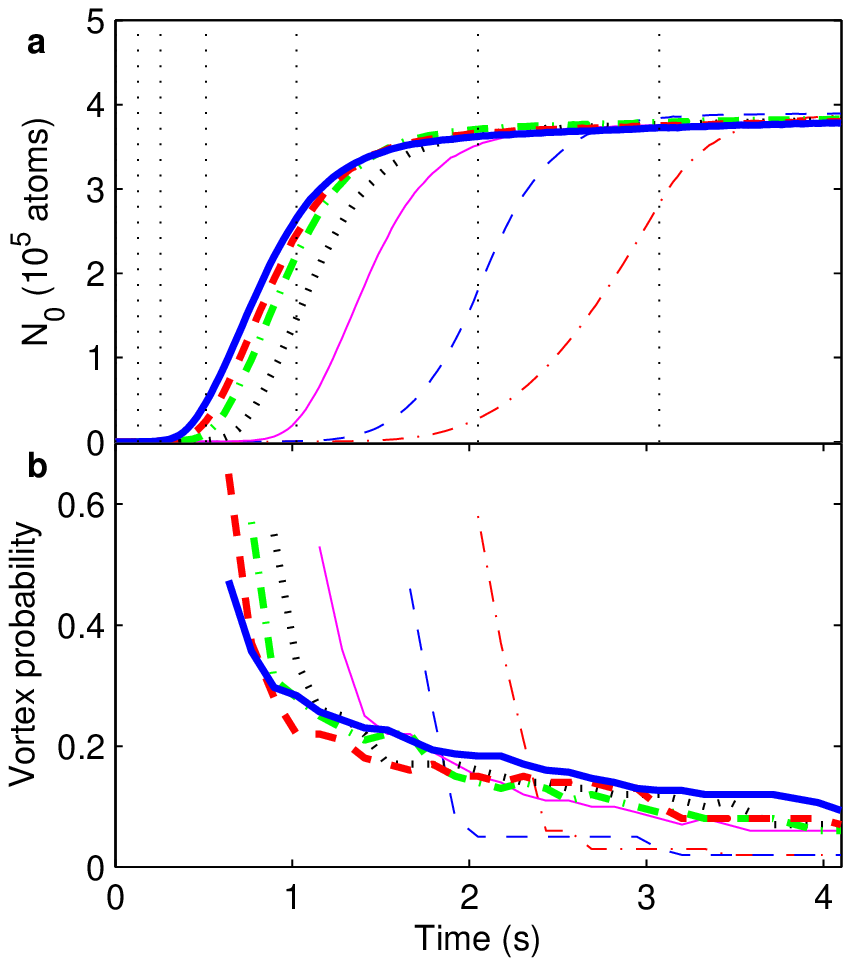}\hspace{2mm}
\includegraphics[width=78mm]{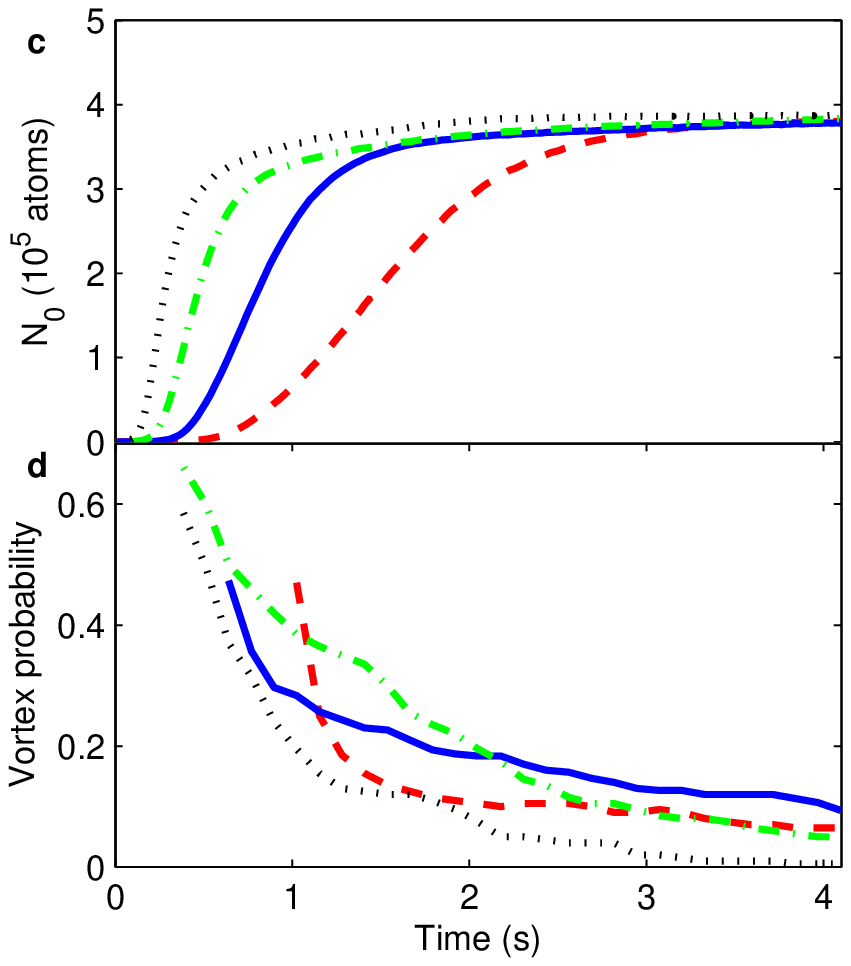}
\caption{\textbf{Effect of altering simulation parameters.}  \textbf{a,b}, Results for condensate growth and vortex probability for a linear ramp of the chemical potential over time scales of $0$ (thick solid blue line), $1$ (thick red dashed line), $2$ (thick green dot-dashed line), $4$ (thick black dotted line), $8$ (thin magenta solid line), $16$ (thin blue dashed line), and $24$ (thin red dot-dashed line) radial trap periods.  The ramp begins at $t=0$, and ends at the times indicated by the vertical black dotted lines.
\textbf{c, d}, Condensate growth and vortex probability for different scattering rates: $\gamma = 0.0025$ (red dashed line), $\gamma = 0.005$ (blue solid line),
$\gamma = 0.01$ (green dot-dash line), and $\gamma = 0.02$ (black dotted line).
}
\label{SIfig2}
\end{figure}

This brings into question the relation between the KZ mechanism of formation of topological defects and the simulations.  In the KZ mechanism, the quench time $\tau_Q$ is a crucial parameter for estimating the initial density of spontaneously formed vortices, and it describes the rate at which the temperature $T$ is lowered relative to the critical temperature $T_c$.  However, given that the condensate forms at the same rate even for different $\mu$ and $T$ ramp times in our simulations (and hence the condensate ``quench'' is occurring on a similar time scale), similar vortex observation probabilities among these simulations seems reasonable.

For the longer time scale ramps of $\mu$ and $T$ over  $16$ and $24$ radial trap periods, the initial sections of the condensate growth curves are much more rounded, and the maximum growth rate is reduced.  Fewer vortices are observed after the condensate reached its final population, and it thus seems that over these time scales the ``effective quench time'' of the system is increased.

The other fitting parameter for the simulations was the dimensionless coupling of the classical field to the thermal cloud, $\gamma = 0.005$, which is essentially a measure of the collision rate within the thermal cloud.  As a check of our methods, we have generated data sets of 200 trajectories for $\gamma = 0.0025, 0.01,$ and $0.02$ with all other parameters unchanged.  Doubling $\gamma$ essentially halves the time scale for condensate formation as shown in Supplementary Figure~\ref{SIfig2}c, and so might lead one to predict that larger $\gamma$ would lead to a higher probability of vortex formation and observation. However, Supplementary Figure~\ref{SIfig2}d shows that while this is sometimes confirmed, the results are complicated by the fact that a larger $\gamma$ also cause vortices to disappear at a faster rate.

To make a definitive statement about the validity of the Kibble-Zurek scenario
for trapped Bose gases, it will be necessary to better control the quenching of
the thermal cloud relative to the condensate.  We are currently working towards
techniques to achieve this.\\%[5mm]

%REFERENCES 

\end{document}